\documentclass[a4paper,fleqn,usenatbib]{mnras}

\usepackage{newtxtext,newtxmath}

\usepackage{pdflscape}

\usepackage[T1]{fontenc}
\usepackage{ae,aecompl}

\usepackage{graphicx}	
\usepackage{amsmath}	
\usepackage{amssymb}	

\newcommand{\fermi}{{\it Fermi}}
\newcommand{\gr}{$\gamma$-ray}
\newcommand{\arcdeg}{$^{\circ}$}

\title[A compact X-ray emitting binary]{A Compact X-Ray Emitting Binary in Likely Association with 4FGL J0935.3+0901}

\author[Z. Wang et al.]{
Zhongxiang Wang,$^{1}$\thanks{E-mail: wangzx@shao.ac.cn}
Yi Xing,$^{1}$
Jujia Zhang,$^{2,3}$
Konstantina Boutsia,$^{4}$
Gege Wang,$^{1}$
\newauthor
V. Jithesh,$^{5}$
Kevin B. Burdge,$^{6}$
Michael W. Coughlin,$^{7}$
Dmitry A. Duev,$^{7}$
\newauthor
S. R. Kulkarni,$^{7,6}$
Reed Riddle$^{6}$
and Eugene Serabyn$^{8}$
\\
$^{1}$Shanghai Astronomical Observatory, Chinese Academy of Sciences, 80 Nandan Road, Shanghai 200030, China\\
$^{2}$Yunnan Observatories, Chinese Academy of Sciences, Kunming 650216, China\\
$^{3}$Key Laboratory for the Structure and Evolution of Celestial Objects, Chinese Academy of Sciences, Kunming 650216, China\\
$^{4}$Las Campanas Observatory, Carnegie Observatories, Casilla 601, La Serena, Chile\\
$^{5}$Inter-University Centre for Astronomy and Astrophysics, Pune 411 007, India\\
$^{6}$Caltech Optical Observatories, California Institute of Technology, Pasadena, CA 91125, USA\\
$^{7}$Division of Physics, Math, and Astronomy, California Institute of Technology, Pasadena, CA 91125, USA\\
$^{8}$Jet Propulsion Laboratory, California Institute of Technology, Pasadena, CA 91109, USA
}

\date{Accepted XXX. Received YYY; in original form ZZZ}

\pubyear{2019}

\begin{document}
\label{firstpage}
\pagerange{\pageref{firstpage}--\pageref{lastpage}}
\maketitle

\begin{abstract}
4FGL J0935.3+0901 is a \gr\ source detected by the Large Area Telescope (LAT)
onboard the {\it Fermi Gamma-Ray Space Telescope}. We have conducted detailed
analysis of the LAT data for this source and multi-wavelength studies of
the source field. Its \gr\ emission can be described with a power law
($\Gamma=2.0\pm0.2$) with an exponential cutoff ($E_c=2.9\pm1.6$ GeV), while
the flux shows significant long-term variations.  
From analysis of archival {\it Neil Gehrels Swift} X-ray Telescope data, 
we find only one X-ray source in the LAT's 2$\sigma$ error region.
Within a 3\farcs7 radius error circle of the X-ray source, 
there is only one 
optical object down to $r'\sim$23 mag. Time-resolved photometry of
the optical object indicates a likely 2.5~hr periodic
modulation, while its spectrum shows double-peaked hydrogen and helium emission
lines (similar to those seen in accretion discs in low-mass X-ray 
binaries). Combining
these results, we conclude that we have discovered a compact X-ray emitting 
binary in likely association with 4FGL J0935.3+0901, i.e., a millisecond
pulsar (MSP) binary.
We discuss the implication of the optical spectral features: this 
binary could be a transitional MSP system at a sub-luminous disc
state, although the other 
possibility, the binary in a rotation-powered state showing the optical 
emission lines due to intrabinary interaction processes, can not be 
excluded.  Further observational studies will help 
determine detailed properties of this candidate MSP binary and 
thus clarify its current state.

\end{abstract}




\section{Introduction}

The unprecedented capabilities of the Large Area Telescope
(LAT) onboard {\it the Fermi Gamma-ray Space Telescope (Fermi)} \citep{atw+09}
have greatly improved our knowledge of high-energy sources in the sky. 
Thus far, more than 200 \gr\ pulsars have been found,\footnote{\footnotesize https://confluence.slac.stanford.edu/display/GLAMCOG/\\Public+List+of+LAT-Detected+Gamma-Ray+Pulsars} 
which has established pulsars as the dominant class of \gr\ sources in our 
Galaxy. Nearly half of them are millisecond pulsars (MSPs). They are
old neutron stars spun up due to the recycling scenario, i.e., their progenitors
are believed to be low-mass X-ray binaries, in which a neutron star gains
angular momentum by accreting mass transferred from a Roche-lobe filling,
low-mass companion star
(e.g., \citealt{alp+82,bv91}). Mainly due to \fermi\ observations, one new 
type of MSP binaries, the redbacks, has been identified \citep{rob13}.
These redbacks 
contain a 
$\ga 0.1\ M_{\sun}$ companion, differing from the long-known black widow 
type of MSP binaries \citep{fst88} in that the latter have very low-mass 
companions ($< 0.1\ M_{\sun}$). Both redbacks and black widows are systems
having the pulsar strongly ablate the low-mass companion, probably mainly
via the pulsar wind. 
Currently more than 20 redbacks or candidates (in which the compact
star has not been confirmed as a pulsar) have been found 
(see \citealt{str+19} and references therein).
These redbacks or candidates were mostly first detected in
\fermi\ LAT observations, and then were identified as compact binaries
from mutli-wavelength observations.

Intriguingly in redbacks, a companion star may 
overfill its Roche lobe,
providing matter to form an accretion disc around the MSP. Currently
there are two field redbacks, PSR J1023+0038 \citep{arc+09}
and XSS J12270$-$4859 \citep{bas+14}, known to be able to have such a 
transition between the states of being a rotation-powered radio binary
MSP (disc-free) and having an accretion disc but with X-ray luminosities
$\la 10^{34}$\,erg\,s$^{-1}$ (i.e., a so-called sub-luminous disc state;
\citealt{lin14,str+19}).
During the latter disc state, one distinguishing feature of redbacks is 
their enhanced \gr\ emission (compared to that in the disc-free state; 
e.g., \citealt{sta+14,xw-XSS}), since generally
low-mass X-ray binaries do not have detectable \gr\ emission. In addition,
a distinctive bimodal count-rate distribution at X-rays is seen in the disc 
state, with the switching timescales between the two count-rate levels as 
short as tens of seconds (e.g., \citealt{pat+14}). Based on these features,
several candidate redbacks have been suggested as transitional MSP systems
(e.g., 3FGL J1544.6$-$1125, \citealt{bh15}; 3FGL J0427.9$-$6704, 
\citealt{slc+16}; FL8Y J1109.8$-$6500, \citealt{cot+19}), 
although no state transition has been seen in them so far.

Identification of redbacks (as well as black widows) not only provides
us with knowledge of the endpoints of binary evolution, helping our
understanding of evolution processes in the phase of low-mass X-ray binaries
(e.g., \citealt{che+13,ben+14}), but also allow to probe the interaction
of a pulsar with its companion via the pulsar wind from multi-wavelength
studies (e.g.,\citealt{bog+11,tak+14,rs16}). Particularly what physical 
properties
drive the transition, i.e., forming a transitional system, are not clear.
Identification of a sample of redbacks as complete as possible will certainly 
help.
In this paper we report our discovery of a likely 2.5-hr binary in the field
of a \fermi\ LAT source. Its optical spectrum shows emission features similar
to those seen in accretion discs in X-ray binaries. 
Given the overall properties (particularly the optical spectral features),
this binary is likely a candidate MSP binary and could even be a
transitional MSP system at its sub-luminous disc state.

\section{Multi-wavelength Data and Analysis Results}
\label{sec:obs}

\subsection{4FGL J0935.3+0901 and LAT Data Analysis}

\begin{figure}
\centering
\includegraphics[width=0.43\textwidth]{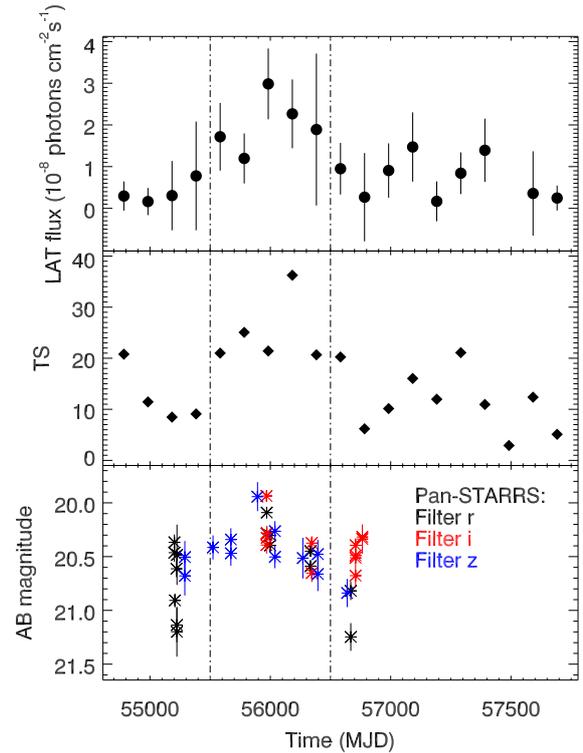}
\caption{{\it Top} and {\it middle panels:} 200-day binned \gr\ light 
curve (0.1--500 GeV) and TS curve
for J0935. Flux data points with TS greater than 5 are kept. Two vertical
dashed lines approximately define the high-flux time period (MJD 55500--56500).
{\it Bottom panel:} Pan-STARRS multi-band magnitude measurements of the optical
binary in the field of J0935 (see Section~\ref{subsec:img}).}
\label{fig:lc}
\end{figure}

The source J0935.3+0901 (hereafter J0935) is an unassociated source recorded 
in the \fermi\ LAT source
catalogs (e.g., \citealt{3fgl15,4fgl19}). Its variable emission 
(Figure~\ref{fig:lc}), revealed
by \gr\ light curve analysis, attracted our attention. 
It actually is newly listed as a variable source in the \fermi\ LAT fourth 
source catalog (4FGL; \citealt{4fgl19}).  We analyzed the LAT data for J0935 
in detail.

LAT events in the 0.1--500 GeV band from the updated \textit{Fermi} Pass 8 
database (P8R3) from 2008-08-04 15:43:36 UTC to 2018-12-02 08:29:55 UTC
were selected. A region of interest (RoI) with a size of
$\mathrm{20^{o}\times20^{o}}$ was selected for analysis, centered at 
the source's 4FGL position
R.A.=143\fdg8326 and Decl.=$+$9\fdg0305 (equinox J2000.0; with a 2$\sigma$ 
nominal uncertainty of 0\fdg07).
We included the LAT events with zenith angles less than 90\arcdeg\ to prevent 
the Earth's limb contamination, and excluded the events with quality flags 
of `bad'. 

For the standard binned likelihood analysis performed to the LAT data,
the source model was based on 4FGL, containing all the 4FGL sources within 
the RoI.
The spectral parameters of the sources within 5\arcdeg\ from J0935
were set free, and those of the remaining sources were fixed to the
catalog values. The background Galactic and extragalactic diffuse 
emission models included were gll\_iem\_v07.fits and 
iso\_P8R3\_SOURCE\_V2\_v1.txt, respectively. The normalizations of the diffuse 
components were set as free parameters.

\gr\ emission of J0935 is described with a simple power law 
$dN/dE = N_{0}E^{-\Gamma}$ in 4FGL. In our analysis using the power law, 
we obtained photon 
index $\Gamma= 2.50\pm$0.07 and 0.1--500 GeV photon flux 
$F_{0.1-500}= 12\pm1\times 10^{-9}$ photons~s$^{-1}$\,cm$^{-2}$, with a 
Test Statistic (TS) value of 233.
We also considered an exponentially cutoff power law 
$dN/dE = N_{0}E^{-\Gamma}\exp(-E/E_{c})$ for J0935, 
where $E_{c}$ is the cutoff energy. This spectral model is characteristic of 
\gr\ emission of pulsars.  We obtained $\Gamma= 2.0\pm$0.2, 
$E_{c}= 2.9\pm$1.6\,GeV, and 
$F_{0.1-500}= 9\pm 2\times 10^{-9}$ photons~s$^{-1}$\,cm$^{-2}$,
with a TS value of 237.
The significance of the spectral cutoff can be estimated from 
$\sqrt{-2\log(L_{pl}/L_{exp})}$, where $L_{exp}$ and $L_{pl}$ are 
the maximum likelihood values obtained from a power law with and without 
the cutoff respectively \citep{2fpsr13}.
For J0935, we found that the spectral cutoff is $\sim$3$\sigma$ significant.
\begin{figure}
\centering
\includegraphics[width=0.45\textwidth]{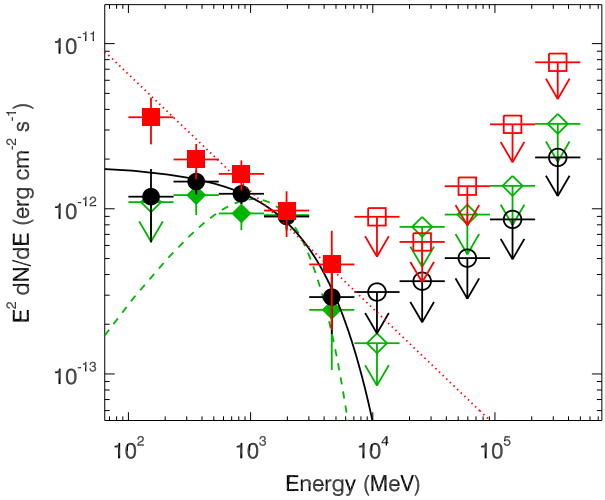}
\caption{\gr\ spectra of J0935 in the total time period (black circles), 
the high-flux time period (red squares), and the low-flux time period (green
diamonds),
with their model fits obtained from the likelihood analysis plotted as 
black solid, red dotted, and green dashed curves, respectively.
Open symbols are flux upper limits.  }
\label{fig:spectra}
\end{figure}

A 200-day binned light curve of J0935 in 0.1--500 GeV was obtained
by performing the maximum likelihood analysis to the data during each  
200-day time intervals. A point source with exponentially cutoff
power-law emission was considered.
In Figure~\ref{fig:lc}, the light curve is shown. Only data points with 
TS greater than 5 ($>$2$\sigma$ significance) were kept in the light curve.
The source's fluxes appear to have 
a factor of 3 variation, although with large uncertainties.
To verify the flux variations, we chose MJD 55500--56500 
as a high-flux time period and the other time ranges as a low-flux time
period, and performed likelihood analysis to the data in the two time periods
respectively.
We found a power law with no cutoff better describing
the emission in the high-flux time period, $\Gamma= 2.7\pm$0.1 and 
$F_{0.1-500}= 24\pm 4\times 10^{-9}$ photons~s$^{-1}$\,cm$^{-2}$
(with a TS value of 121). In the low-flux time period, 
the spectral cutoff was significant ($\sim$5$\sigma$),
$\Gamma= 1.0\pm$0.5, $E_{c}= 1.1\pm$0.4 GeV, and 
$F_{0.1-500}= 3\pm 1\times 10^{-9}$ photons~s$^{-1}$\,cm$^{-2}$ 
(with a TS value of 107). Therefore, J0935 not only showed approximately 
a factor of 8 flux variation, but also had related spectral 
changes.

We extracted the \gr\ spectra of J0935 in the total time period as well as
in the high-flux and low-flux time periods.
Maximum likelihood analysis was performed to the LAT data in 
10 evenly divided energy bands in logarithm from 0.1 GeV to 500 GeV. 
The spectral normalizations of the sources within 5\arcdeg\ from J0935 were
set as free parameters, and all the other parameters of the sources in 
the source model were fixed at the values obtained from the above maximum 
likelihood analysis.  The obtained spectra are shown in 
Figure~\ref{fig:spectra}, and the fluxes and TS values are given 
in Table~\ref{tab:spectra}, in which we kept only spectral data points 
with TS values greater than 5 ($>$2$\sigma$ significance), and 
provided 95\% (2$\sigma$) flux upper limits otherwise. 

\begin{figure}
\centering
\includegraphics[width=0.37\textwidth]{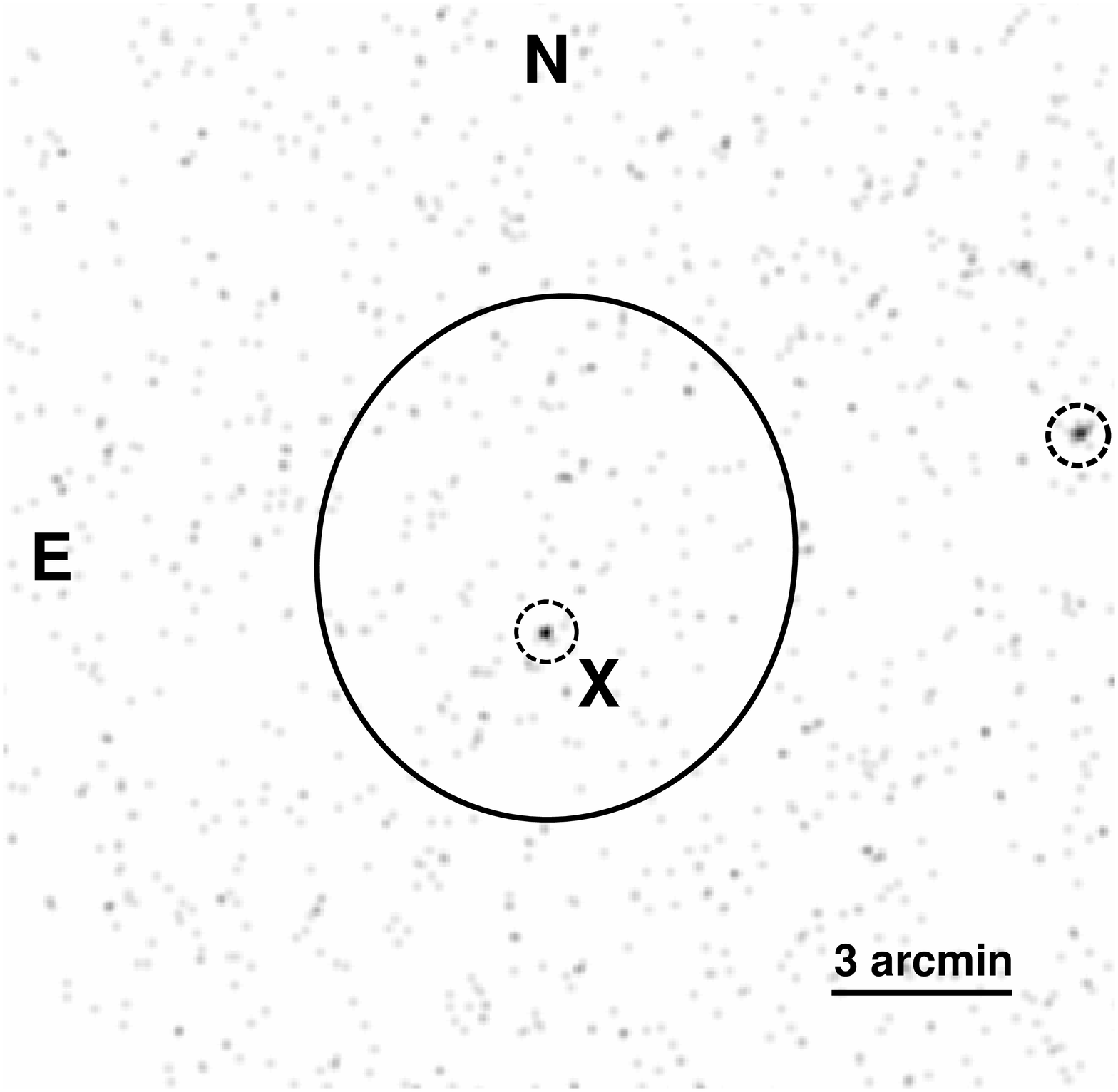}
\includegraphics[width=0.37\textwidth]{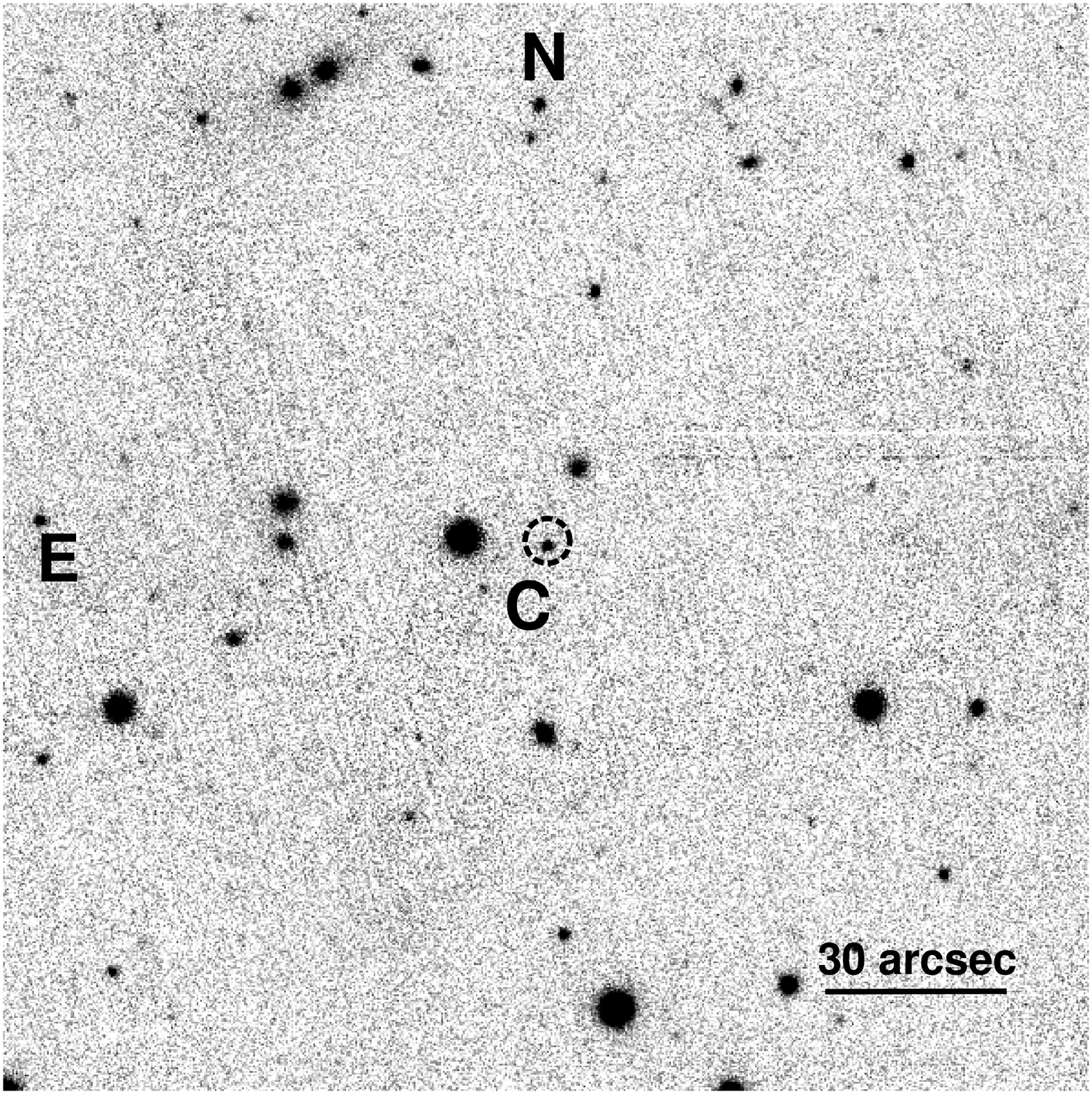}
\caption{X-ray and optical $r'$ band images ({\it top} and {\it bottom} panels,
respectively) of the J0935 field. Two X-ray sources were detected in
the field (marked by dashed circles), and one of them (object $X$) is
in the 2$\sigma$ LAT error region (large solid ellipse). 
In the X-ray error circle (3\farcs7 radius) of object $X$, there is
only one
optical source (object $C$) detected by Pan-STARRS down to $r'\sim 21$ mag.}
\label{fig:finding}
\end{figure}
\subsection{Analysis of {\it Swift} X-Ray Telescope Data}

Since positional uncertainties from \fermi\ LAT observations are relatively 
large (for which a 2$\sigma$ uncertainty region is mostly secure enough to
be searched for a counterpart), we searched archival X-ray data that cover the
field of J0935 for possible counterpart identification. 
The {\it Neil Gehrels Swift Observatory} \citep{swift+04} performed 11 short (0.1--1.0\,ks) 
observations of the source field between 2015 Jan 29 
and 2016 Jun  26. We analyzed all the data taken with 
the X-ray telescope (XRT; \citealt{xrt+05}) and the total
exposure is 3.64 ks.
The XRT data were reduced and re-processed by
applying the standard filtering and screening criteria.
The {\sc xrtpipeline} tool version~0.13.4 in the {\sc heasoft} package 
version~6.24 was used. We collected the photon counting (PC) mode data and 
used a 
sliding-cell detection algorithm in {\sc ximage} to detect sources. 
No sources were detected in any individual observations at a $3\sigma$ 
significance level. We added all the observations and were able to
detect two sources in the combined image 
(top panel of Figure~\ref{fig:finding}). Only one of them
(object $X$) was in the \fermi\ LAT 2$\sigma$ error region. 
Using the {\sc xrtcentroid} tool, we derived the source's position,
RA = $9^{h}35^{m}20\fs60$, Dec. = $+9^{\circ}00^{'}35\farcs4$ 
(equinox J2000.0), with a positional uncertainty of 5\farcs5 (at a
90\% confidence level). The position was further improved using the online
tool provided by the UK {\it Swift} Science Data Centre\footnote{\footnotesize https://www.swift.ac.uk/user\_objects/}. 
The resulting enhanced position \citep{goa+07,eva+09}
is RA = $9^{h}35^{m}20\fs70$, Dec. = $+9^{\circ}00^{'}36\farcs8$
(equinox J2000.0), with a smaller uncertainty of 3\farcs7 (at a
90\% confidence level).

Source and background events were extracted from a circular region with a 
radius of 24\arcsec, and the spectra of the source and background were
generated from the cleaned event files with the standard grade filtering 
of 0--12. The ancillary response files were also generated, using 
the tool {\sc xrtmkarf} and the spectral redistribution matrices available 
in the calibration database (CALDB version 20180710). 
The extracted source spectrum has limited net counts (14 counts in total).
We used the Cash Statistics \citep{cas79} for the model fitting. 
An absorbed power-law model in the $0.3-10$~keV energy band was used,
with the hydrogen column density fixed at the Galactic value 
$3.6\times 10^{20}\ \rm cm^{-2}$ \citep{hi+16}. 
The model fit did not yield well-constrained results: power-law index is 
$2.3^{+1.5}_{-1.4}$ and the unabsorbed flux is 
$(1.2^{+1.6}_{-0.7})\times 10^{-13}\ \rm erg~cm^{-2}~s^{-1}$ 
(cstat=8.1 with 11 degrees of freedom).

We also searched for fast count-rate variations by 
generating X-ray light curves of object $X$ with different short time bins, 
although the source was faint and the {\it Swift} observations are very short. 
No clear variations were seen. In addition, we checked the 70- and 105-month
{\it Swift} Burst Alert Telescope catalogs \citep{bat70,bat105}, and the source 
was not detected. The flux upper limit on the hard X-ray emission from the
source in the 14--195~keV band thus was 
$\sim 10^{-11}$\,erg\,s$^{-1}$\,cm$^{-2}$.

\begin{figure}
\centering
\includegraphics[width=0.45\textwidth]{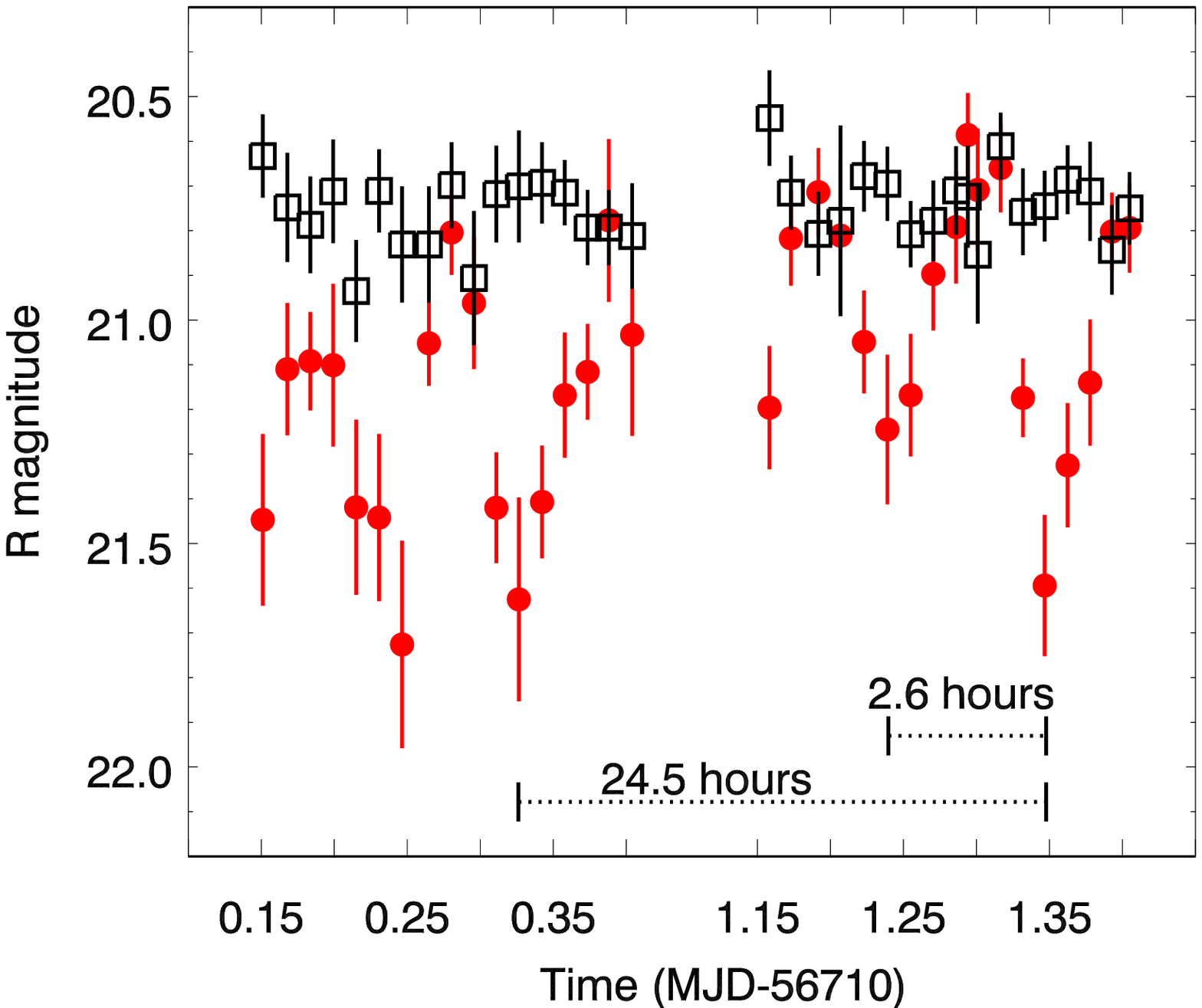}
\includegraphics[width=0.45\textwidth]{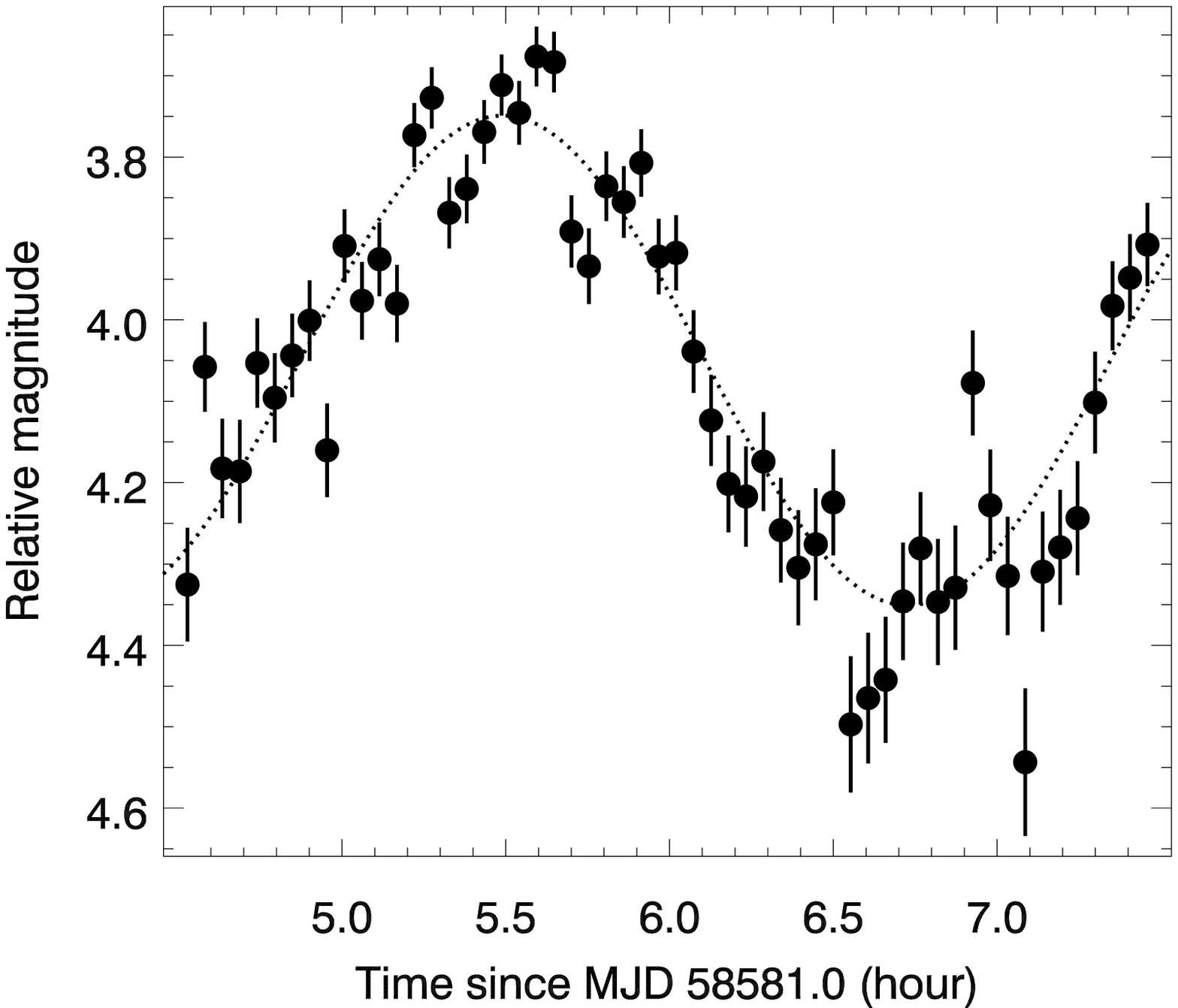}
\caption{{\it Top panel:} PTF $R$ band light curve of object $C$
(red circles). 
A possible 2.5 hr periodic modulation is seen, based on the modulation
dips.
For comparison, the light curve of a check star (black squares) 
is also shown.
{\it Bottom panel:} KPED $r'$ light curve of object $C$, showing
sinusoidal-like modulation in detail. A sinusoidal function with a period of
2.47$\pm$0.04 hrs and a semi-amplitude of 0.30$\pm$0.01 mag (dotted curve)
can describe the modulation.}
\label{fig:olc}
\end{figure}

\subsection{Optical Observations and Data Analysis}
\subsubsection{Imaging}
\label{subsec:img}
The Panoramic Survey Telescope and Rapid Response System (Pan-STARRS; 
\citealt{pans16,pand16}) observed the field of J0935 multiple times.
Only one optical object (object $C$; bottom panel of 
Figure~\ref{fig:finding}), which was variable and had magnitudes in a range of
20--21.5 (top panel of Figure~\ref{fig:lc}), was found in 
the {\it Swift} X-ray error circle. We note that 
from the Pan-STARRS measurements, object $C$ showed a brightening possibly
correlated with the high-flux time period of J0935. In addition, the Sloan
Digital Sky Survey (SDSS) observed the source field once in 2003 Jan. 
Object~$C$ was still the only source detected in the X-ray error circle and
its multi-band magnitudes are in line with the Pan-STARRS measurements
(e.g., its $r'=20.433\pm0.044$ in SDSS; \citealt{aba+09}). The SDSS imaging 
has an upper limit of $r'\sim 23$~mag (3$\sigma$) on the source field,
which implies that there were no other objects in the X-ray error circle
down to $r'\sim 23$~mag.

We also checked the archival data from the Palomar Transient Factory 
(PTF; \citealt{ptf09}). The source field was well observed on 2014 Feb 22--23
in $R$ band, with an exposure time of 60\,s per frame. For the image data,
we combined every three consecutive frames to
increase the detection significance of object $C$. In total,
35 images with good point-spread functions (PSFs) of sources
over the two nights were obtained. We performed 
standard PSF-fitting photometry to the object and obtained
its light curve, in which 9 relatively bright stars in the field without
obvious variations were used for relative flux calibrations among the images.
The same photometry was performed on a check star, which 
was chosen given its brightness similar to object $C$.
The light curves of object $C$ and the check star are shown 
in Figure~\ref{fig:olc}. Comparing to the check star, significant modulation 
is seen in object $C$. Based on the times of the modulation dips,
a likely period of $\sim 2.5$ hrs was found.

To confirm the periodic modulation of object $C$, we further
obtained
a $r'$ light curve of it with the Kitt Peak electron multiplying CCD 
demonstrator (KPED; \citealt{kped19}) on 2019 Apr 8. The total 
observation time was three hours. The exposure time per frame was
30\,s. KPED aperture
photometry pipeline was used to measure the brightness of the 
object. The final light curve, shown in Figure~\ref{fig:olc}, was re-binned 
from the original light curve by 
averaging every 6 measurements. A sinusoidal-like modulation
is seen in the light curve. We fit the light curve with a sinusoidal
function, and found a period of 2.47$\pm$0.04\,hrs
and a semi-amplitude of 0.30$\pm$0.01 mag ($\chi^2$ is 137 with 52
degrees of freedom; see Figure~\ref{fig:olc}). Several data points near
the bottom of the modulation appear to deviate away from the sinusoidal fit,
which should be verified from more sensitive photometry in order to check 
if it is a true feature. 

Sinusoidal-like modulation is commonly seen in low-mass X-ray
binaries or pulsar binaries (see, e.g., \citealt{vm95,str+19}). 
The inner face of the companion star is heated by X-ray emission or pulsar 
wind from the central compact object. The visible area of the heated face
varies as a function of orbital phase, giving rise to the sinusoidal 
modulation in these binaries. The observed optical modulation thus likely
indicates that object $C$ is a 2.5 hr binary. However to be
cautious, because our
KPED observation only lasted 3 hrs, we can not totally exclude the possibility
of the light curve being ellipsoidal. Such modulation has two 
maxima and two
minima over one orbit and is caused by the orbital motion of the companion
star when there is no strong irradiation and the star is tidally distorted
(e.g., \citealt{vm95}; also see \citealt{yap+19} for an interesting redback 
case changing from ellipsoidal to sinusoidal-like modulation). If this is 
the case, the orbital period would be $\sim$5 hrs.
\begin{figure}
\centering
\includegraphics[width=0.47\textwidth]{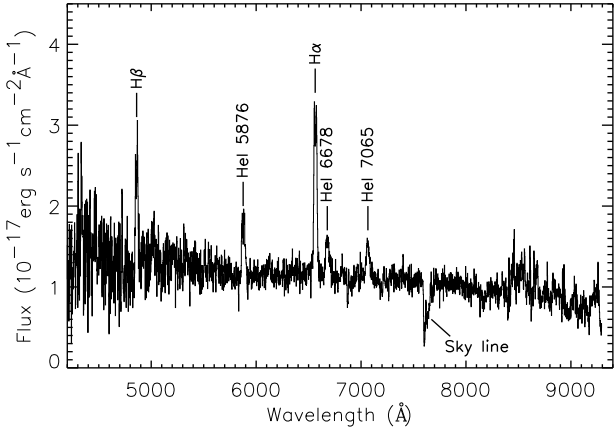}
\caption{Magellan LDSS3 spectrum of object $C$. The detected
H$\alpha$, H$\beta$, and three He I emission lines are marked.
}
\label{fig:spec}
\end{figure}

\subsubsection{Magellan Spectroscopy}

On 2019 Feb 21, we obtained an optical spectrum of object $C$.
The observation was conducted with the Low Dispersion Survey Spectrograph 
(LDSS3) on 6.5-m Magellan II (Clay) telescope.
We chose the VPH-ALL grism with a long slit of 0.75\arcsec\ in width,
covering 4250--10000\,\AA, and providing a dispersion of 1.89 \AA\,pixel$^{-1}$.

Three 15-min exposures of the source were taken. 
LTT4364, a spectrophotometric standard star, was observed for flux calibration.
The three reduced spectral images were combined to yield a 
45-min total exposure spectrum image. The source spectrum was extracted
using {\tt IRAF} spectral reduction and analysis tasks. The spectrum
(Figure~\ref{fig:spec})
shows prominent double-peaked H$\alpha$, H$\beta$, and three
\ion{He}{i} emission lines. These emission features are typically seen
in spectra of accretion discs in X-ray binaries and also similar
to that of, for example, PSR J1023+0038 in the disc state 
(see \citealt{wan+09}). Using two Gaussian functions to fit
the profiles of H$\alpha$ and \ion{He}{i} $\lambda$5876 lines, which have
clear double-peak structures, we estimated their properties.
For H$\alpha$, the blue and red peaks have velocities
of $-$430$\pm82$ and 402$\pm96$\,km\,s$^{-1}$ (with FWHMs of 
463$\pm$90 and 575$\pm$150\,km\,s$^{-1}$) respectively,
and for \ion{He}{i} $\lambda$5876, the values are $-587\pm174$
and 403$\pm276$\,km\,s$^{-1}$ (FWHMs are approximately
378 and 760\,km\,s$^{-1}$, with $\sim$50\% uncertainties). 
For comparison, these values 
are similar to those of PSR J1023+0038 in the disc state 
obtained in \citet{wan+09}.

Since the optical counterpart is faint (to the Magellan telescope),
the blue end of the spectrum suffers large uncertainties, but we 
suspect that there are other hydrogen and helium emission features. 
Also no obvious, identifiable absorption features are seen in the spectrum. 
These should be investigated through spectroscopy on a larger telescope.
In order to confirm the likely orbital period of 2.5~hrs,
phase-resolved spectroscopy of absorption lines of the
companion star (to obtain its radial velocity curve)
is needed.

\section{Discussion}

Having analyzed archival X-ray and optical data as well as optical data
obtained in our follow-up observations, we have found a binary with
likely 2.5-hr orbital period in the error region of the \fermi\ LAT
source J0935. The optical spectrum of the binary 
contains double-peaked hydrogen and helium emission lines, which are
typically seen in
accretion discs in X-ray binaries. Given the small error circle 
of the X-ray source (with a radius of 3\farcs7 
at a 90\% confidence level) and the binary
is the only optical object in the error circle down to $r'\sim 23$~mag, 
it can be concluded that we have discovered a compact X-ray emitting binary.
While normal compact X-ray binaries, which for example contain an
accreting neutron star, are not \gr\ sources,
redbacks have both observable X-ray and \gr\ emission and particularly
the transitional MSP systems in the sub-luminous disc state have 
enhanced \gr\ emission. Thus whether the 2.5-hr binary is
associated with J0935 determines the nature of this binary.

At high Galactic latitudes ($b\sim$40\arcdeg\ for J0935), the dominant 
\gr\ sources are blazars, and their spectra are of power-law, log-parabola, 
or broken power-law shapes (e.g., \citealt{3fagn15}).
For J0935, we have shown that it has a curved \gr\ spectrum, 
a $\Gamma\simeq 2.0$ power law with exponential cutoff $E_c \sim 2.9$\,GeV.
Such a spectral form is characteristic of pulsars' \gr\ emission 
(e.g., \citealt{2fpsr13,xw16}). 
The \gr\ spectrum is indeed similar to those of PSR J1023$+$0038 and 
XSS J12270$-$4859 in the disc state \citep{xwt18,xw-XSS}, having 
$\Gamma\sim 2$ and several GeV cutoff. 
The currently known (candidate) redbacks were mostly
discovered from searches for binaries in the fields of candidate MSP systems,
identified from \fermi\ LAT data analysis (\citealt{str+19} and references 
therein). Also their \gr\ luminosities are generally in a range of 
$\sim$10$^{33}$--5$\times 10^{34}$\,erg\,s$^{-1}$, with X-ray counterparts 
having luminosities of $\sim 10^{31}$--$10^{34}$\,erg\,s$^{-1}$. It can be
noted that the (candidate) transitional systems at the disc state are mostly
at the high end of the ranges.  While the LAT positional uncertainties are
relatively large, we found only one X-ray source, the 2.5-hr binary, in
the $\simeq 0\fdg07$ (2$\sigma$) error region of J0935 down to
$\sim 10^{-13}$\,erg\,cm$^{-2}$\,s$^{-1}$.
Therefore on the basis of the \gr\ spectral properties of J0935
and the X-ray detection in the LAT error region expected for a redback, 
we suggest that J0935 is associated with the 2.5-hr binary.
In addition, we note that a possible correlated optical brightening of
the binary was seen, 
when the long-term \gr\ light curve of J0935 was in a high-flux time
period (Figure~\ref{fig:lc}).

One notable feature of J0935 is its \gr\ flux variations, which
was determined by comparing the fluxes in the two time periods we defined.
This feature has not been determined to be common for redbacks or the two 
transitional systems. 
In the latter two systems, significant flux changes occurred over their 
state transition by appearing ``step-like"
(e.g., \citealt{sta+14,xw-XSS,tor+17}), while the same spectral form 
was kept. In \citet{tor+17}, \gr\ flux variations of a few redbacks 
(including the two transitional systems) and black widows
were searched. XSS J12270$-$4859 in the disc state 
was found to have significant variations and two black widows 
(PSR J1446$-$4701 and PSR J2234+0944) also had variations. We checked and 
found that the \gr\  light curve 
analysis methods used in \citet{tor+17} for their search would not show 
J0935 as a variable (due to large flux uncertainties). In any case,
the \gr\ variability of J0935 helps reinforce its compact-binary nature,
as isolated pulsars generally have non-variable \gr\ emission \citep{2fpsr13}.

The optical spectral features of the binary, the double-peaked emission
lines, would further suggest that this is   
another candidate transitional MSP system at its disc state.
This system would have been in the disc state all the time, since
our optical spectrum was taken recently and J0935 has 
been in its low-flux
period since 2013 August based on the long-term \gr\ light curve.
However it has been noted that several redbacks (and a black widow;
PSR J1740$-$5340A, \citealt{sab+03}; 3FGL~J0838.8$-$2829, \citealt{hsl17};
2FGL~J0846.0+2820, \citealt{swi+18}; PSR J1048+2339, \citealt{str+19};
PSR J1311$-$3430, \citealt{rfc15})
can have strong variable emission lines and the lines (notable the H$\alpha$
line) can appear to be double-peaked. This feature is interpreted as the
indication of material driven from the heated side of a companion by
the pulsar wind and possibly plus an intrabinary shock between the companion
and the pulsar wind (\citealt{str+19} and references therein).
In the sub-luminous disc state, the $\gamma$-ray--to--X-ray flux ratios 
of the transitional systems and candidates are $\sim 1$ \citep{str+19,cot+19}.
For J0935, its average \gr\ flux (in 0.1--500 GeV) is 
$\simeq 4.8\times 10^{-12}$\,erg\,s$^{-1}$\,cm$^{-2}$ and 
the flux ratio is $\sim 40$. Although the X-ray flux is not well determined
with high uncertainty, the face value of the flux ratio is more in line
with those of redbacks in a rotation-powered state \citep{str+19}. Therefore
we can not exclude the possibility that J0935 is a rotation-powered binary
MSP and 
whether or not it is a candidate transitional system remains to be further
investigated. Optical spectroscopy with a large telescope to check
the stableness of the double-peaked emission lines will help clarify this
uncertainty.

In order to confirm J0935 as a MSP binary and determine its possible 
transitional nature,  multi-band phase-resolved photometry
is warranted, which allows to verify the likely 2.5-hr orbital modulation
and moreover to derive detailed properties of the binary
(see \citealt{str+19} and references therein). 
Time-resolved photometry of PSR J1023+0038 in the disc state
has revealed optical variations more complex than simple sinusoidal
modulation, likely reflecting significant activity of the accretion 
disc \citep{sha+15,pap+18,sha+18,ken+18}.
The possible deviation of the KPED light curve (at the bottom of
the modulation; Figure~\ref{fig:olc}) from the sinusoidal fit
might suggest similar variations in J0935. The photometry, while requiring 
a high cadence, will help establish the similarity.
At X-rays, a sensitive observation should be conducted, to 
obtain the X-ray spectrum and check if there are large-amplitude and 
fast flux variations.
Power-law emission with photon index $\sim 1.3$ and 1.6--1.7
is often seen in redbacks in the rotation-powered state and transitional
systems in the disc state respectively,
with certain X-ray luminosity values at the different 
states (e.g., \citealt{lin14,bh15,cot+19}), and 
the transitional systems in the disc state 
show their distinguishing feature of having fast switches between two 
different count-rate levels (e.g., \citealt{pat+14,bh15,cot+19}). Detection
of such variability would certainly help establish J0935
as a candidate transitional MSP system. 
At last \fermi\ LAT observations of J0935 provides monitoring checks.
If J0935 is indeed a candidate transitional system, 
the transition from the current
sub-luminous disc state to a rotation-powered state would be expected. 
Once the \gr\ flux drops significantly in a short time period,
it would likely indicate a transition, which could be verified with 
multi-wavelength observations by searching for pulsed emission from
the MSP and detecting related flux drops at X-rays and optical
bands.

\section*{Acknowledgements}
This paper includes data gathered with the 6.5 meter Magellan Telescopes 
located at Las Campanas Observatory, Chile.
This research made use of the High Performance Computing Resource in the Core
Facility for Advanced Research Computing at Shanghai Astronomical Observatory.
This research was supported by the National Program on Key Research 
and Development Project (Grant No. 2016YFA0400804) and
the National Natural Science Foundation
of China (11633007, U1738131). J. Zhang is supported by the NSFC 
(grants 11773067, 11403096), the Youth Innovation Promotion Association of 
the CAS (grants 2018081), and  the Western Light Youth Project.

\bibliographystyle{mnras}


\begin{thebibliography}{}
\makeatletter
\relax
\def\mn@urlcharsother{\let\do\@makeother \do\$\do\&\do\#\do\^\do\_\do\%\do\~}
\def\mn@doi{\begingroup\mn@urlcharsother \@ifnextchar [ {\mn@doi@}
  {\mn@doi@[]}}
\def\mn@doi@[#1]#2{\def\@tempa{#1}\ifx\@tempa\@empty \href
  {http://dx.doi.org/#2} {doi:#2}\else \href {http://dx.doi.org/#2} {#1}\fi
  \endgroup}
\def\mn@eprint#1#2{\mn@eprint@#1:#2::\@nil}
\def\mn@eprint@arXiv#1{\href {http://arxiv.org/abs/#1} {{\tt arXiv:#1}}}
\def\mn@eprint@dblp#1{\href {http://dblp.uni-trier.de/rec/bibtex/#1.xml}
  {dblp:#1}}
\def\mn@eprint@#1:#2:#3:#4\@nil{\def\@tempa {#1}\def\@tempb {#2}\def\@tempc
  {#3}\ifx \@tempc \@empty \let \@tempc \@tempb \let \@tempb \@tempa \fi \ifx
  \@tempb \@empty \def\@tempb {arXiv}\fi \@ifundefined
  {mn@eprint@\@tempb}{\@tempb:\@tempc}{\expandafter \expandafter \csname
  mn@eprint@\@tempb\endcsname \expandafter{\@tempc}}}

\bibitem[\protect\citeauthoryear{{Abazajian} et~al.,}{{Abazajian}
  et~al.}{2009}]{aba+09}
{Abazajian} K.~N.,  et~al., 2009, \mn@doi [\apjs]
  {10.1088/0067-0049/182/2/543}, \href
  {https://ui.adsabs.harvard.edu/abs/2009ApJS..182..543A} {182, 543}

\bibitem[\protect\citeauthoryear{{Abdo} et~al.,}{{Abdo}
  et~al.}{2013}]{2fpsr13}
{Abdo} A.~A.,  et~al., 2013, \mn@doi [\apjs] {10.1088/0067-0049/208/2/17},
  \href {http://adsabs.harvard.edu/abs/2013ApJS..208...17A} {208, 17}

\bibitem[\protect\citeauthoryear{{Acero} et~al.,}{{Acero}
  et~al.}{2015}]{3fgl15}
{Acero} F.,  et~al., 2015, \mn@doi [\apjs] {10.1088/0067-0049/218/2/23}, \href
  {http://adsabs.harvard.edu/abs/2015ApJS..218...23A} {218, 23}

\bibitem[\protect\citeauthoryear{Ackermann et~al.}{Ackermann
  et~al.}{2015}]{3fagn15}
Ackermann M.,  et~al., 2015, \mn@doi [\apj]
  {10.1088/0004-637X/810/1/14}, 810, 14


\bibitem[\protect\citeauthoryear{{Alpar}, {Cheng}, {Ruderman}  \&
  {Shaham}}{{Alpar} et~al.}{1982}]{alp+82}
{Alpar} M.~A.,  {Cheng} A.~F.,  {Ruderman} M.~A.,   {Shaham} J.,  1982, \mn@doi
  [\nat] {10.1038/300728a0}, \href
  {https://ui.adsabs.harvard.edu/abs/1982Natur.300..728A} {300, 728}

\bibitem[\protect\citeauthoryear{{Archibald} et~al.,}{{Archibald}
  et~al.}{2009}]{arc+09}
{Archibald} A.~M.,  et~al., 2009, \mn@doi [Science] {10.1126/science.1172740},
  \href {http://adsabs.harvard.edu/abs/2009Sci...324.1411A} {324, 1411}

\bibitem[\protect\citeauthoryear{{Atwood} et~al.,}{{Atwood}
  et~al.}{2009}]{atw+09}
{Atwood} W.~B.,  et~al., 2009, \mn@doi [\apj] {10.1088/0004-637X/697/2/1071},
  \href {http://adsabs.harvard.edu/abs/2009ApJ...697.1071A} {697, 1071}

\bibitem[\protect\citeauthoryear{{Bassa} et~al.,}{{Bassa}
  et~al.}{2014}]{bas+14}
{Bassa} C.~G.,  et~al., 2014, \mn@doi [\mnras] {10.1093/mnras/stu708}, \href
  {http://adsabs.harvard.edu/abs/2014MNRAS.441.1825B} {441, 1825}

\bibitem[\protect\citeauthoryear{{Baumgartner}, {Tueller}, {Markwardt},
  {Skinner}, {Barthelmy}, {Mushotzky}, {Evans}  \& {Gehrels}}{{Baumgartner}
  et~al.}{2013}]{bat70}
{Baumgartner} W.~H.,  {Tueller} J.,  {Markwardt} C.~B.,  {Skinner} G.~K.,
  {Barthelmy} S.,  {Mushotzky} R.~F.,  {Evans} P.~A.,   {Gehrels} N.,  2013,
  \mn@doi [\apjs] {10.1088/0067-0049/207/2/19}, \href
  {https://ui.adsabs.harvard.edu/abs/2013ApJS..207...19B} {207, 19}

\bibitem[\protect\citeauthoryear{{Bhattacharya} \& {van den
  Heuvel}}{{Bhattacharya} \& {van den Heuvel}}{1991}]{bv91}
{Bhattacharya} D.,  {van den Heuvel} E.~P.~J.,  1991, \mn@doi [\physrep]
  {10.1016/0370-1573(91)90064-S}, \href
  {https://ui.adsabs.harvard.edu/abs/1991PhR...203....1B} {203, 1}

\bibitem[\protect\citeauthoryear{{Benvenuto}, {De Vito}  \&
  {Horvath}}{{Benvenuto} et~al.}{2014}]{ben+14}
{Benvenuto} O.~G.,  {De Vito} M.~A.,   {Horvath} J.~E.,  2014, \mn@doi [\apjl]
  {10.1088/2041-8205/786/1/L7}, \href
  {http://adsabs.harvard.edu/abs/2014ApJ...786L...7B} {786, L7}

\bibitem[\protect\citeauthoryear{{Bogdanov} \& {Halpern}}{{Bogdanov} \&
  {Halpern}}{2015}]{bh15}
{Bogdanov} S.,  {Halpern} J.~P.,  2015, \mn@doi [\apjl]
  {10.1088/2041-8205/803/2/L27}, \href
  {https://ui.adsabs.harvard.edu/abs/2015ApJ...803L..27B} {803, L27}

\bibitem[\protect\citeauthoryear{{Bogdanov}, {Archibald}, {Hessels}, {Kaspi},
  {Lorimer}, {McLaughlin}, {Ransom}  \& {Stairs}}{{Bogdanov}
  et~al.}{2011}]{bog+11}
{Bogdanov} S.,  {Archibald} A.~M.,  {Hessels} J.~W.~T.,  {Kaspi} V.~M.,
  {Lorimer} D.,  {McLaughlin} M.~A.,  {Ransom} S.~M.,   {Stairs} I.~H.,  2011,
  \mn@doi [\apj] {10.1088/0004-637X/742/2/97}, \href
  {http://adsabs.harvard.edu/abs/2011ApJ...742...97B} {742, 97}

\bibitem[\protect\citeauthoryear{{Burrows} et~al.,}{{Burrows}
  et~al.}{2005}]{xrt+05}
{Burrows} D.~N.,  et~al., 2005, \mn@doi [\ssr] {10.1007/s11214-005-5097-2},
  \href {https://ui.adsabs.harvard.edu/abs/2005SSRv..120..165B} {120, 165}

\bibitem[\protect\citeauthoryear{{Cash}}{{Cash}}{1979}]{cas79}
{Cash} W.,  1979, \mn@doi [\apj] {10.1086/156922}, \href
  {http://adsabs.harvard.edu/abs/1979ApJ...228..939C} {228, 939}

\bibitem[\protect\citeauthoryear{{Chambers} et~al.,}{{Chambers}
  et~al.}{2016}]{pans16}
{Chambers} K.~C.,  et~al., 2016, arXiv e-prints, \href
  {https://ui.adsabs.harvard.edu/abs/2016arXiv161205560C} {p. arXiv:1612.05560}

\bibitem[\protect\citeauthoryear{{Chen}, {Chen}, {Tauris}  \& {Han}}{{Chen}
  et~al.}{2013}]{che+13}
{Chen} H.-L.,  {Chen} X.,  {Tauris} T.~M.,   {Han} Z.,  2013, \mn@doi [\apj]
  {10.1088/0004-637X/775/1/27}, \href
  {http://adsabs.harvard.edu/abs/2013ApJ...775...27C} {775, 27}

\bibitem[\protect\citeauthoryear{{Coti Zelati} et~al.,}{{Coti Zelati}
  et~al.}{2019}]{cot+19}
{Coti Zelati} F.,  et~al., 2019, \mn@doi [\aap] {10.1051/0004-6361/201834835},
  \href {https://ui.adsabs.harvard.edu/abs/2019A&A...622A.211C} {622, A211}

\bibitem[\protect\citeauthoryear{{Coughlin} et~al.,}{{Coughlin}
  et~al.}{2019}]{kped19}
{Coughlin} M.~W.,  et~al., 2019, \mn@doi [\mnras] {10.1093/mnras/stz497}, \href
  {https://ui.adsabs.harvard.edu/abs/2019MNRAS.485.1412C} {485, 1412}


\bibitem[\protect\citeauthoryear{{Evans} et~al.,}{{Evans}
  et~al.}{2009}]{eva+09}
{Evans} P.~A.,  et~al., 2009, \mn@doi [\mnras]
  {10.1111/j.1365-2966.2009.14913.x}, \href
  {https://ui.adsabs.harvard.edu/abs/2009MNRAS.397.1177E} {397, 1177}

\bibitem[\protect\citeauthoryear{{Flewelling} et~al.,}{{Flewelling}
  et~al.}{2016}]{pand16}
{Flewelling} H.~A.,  et~al., 2016, arXiv e-prints, \href
  {https://ui.adsabs.harvard.edu/abs/2016arXiv161205243F} {p. arXiv:1612.05243}

\bibitem[\protect\citeauthoryear{{Fruchter}, {Stinebring}  \&
  {Taylor}}{{Fruchter} et~al.}{1988}]{fst88}
{Fruchter} A.~S.,  {Stinebring} D.~R.,   {Taylor} J.~H.,  1988, \mn@doi [\nat]
  {10.1038/333237a0}, \href {http://adsabs.harvard.edu/abs/1988Natur.333..237F}
  {333, 237}

\bibitem[\protect\citeauthoryear{{Gehrels} et~al.,}{{Gehrels}
  et~al.}{2004}]{swift+04}
{Gehrels} N.,  et~al., 2004, \mn@doi [\apj] {10.1086/422091}, \href
  {https://ui.adsabs.harvard.edu/abs/2004ApJ...611.1005G} {611, 1005}

\bibitem[\protect\citeauthoryear{{Goad} et~al.,}{{Goad} et~al.}{2007}]{goa+07}
{Goad} M.~R.,  et~al., 2007, \mn@doi [\aap] {10.1051/0004-6361:20078436}, \href
  {https://ui.adsabs.harvard.edu/abs/2007A&A...476.1401G} {476, 1401}

\bibitem[\protect\citeauthoryear{{HI4PI Collaboration} et~al.,}{{HI4PI
  Collaboration} et~al.}{2016}]{hi+16}
{HI4PI Collaboration} et~al., 2016, \mn@doi [\aap]
  {10.1051/0004-6361/201629178}, \href
  {https://ui.adsabs.harvard.edu/abs/2016A%26A...594A.116H} {594, A116}

\bibitem[\protect\citeauthoryear{{Halpern}, {Strader}  \& {Li}}{{Halpern}
  et~al.}{2017}]{hsl17}
{Halpern} J.~P.,  {Strader} J.,   {Li} M.,  2017, \mn@doi [\apj]
  {10.3847/1538-4357/aa7cff}, \href
  {https://ui.adsabs.harvard.edu/abs/2017ApJ...844..150H} {844, 150}


\bibitem[\protect\citeauthoryear{{Kennedy}, {Clark}, {Voisin}  \&
  {Breton}}{{Kennedy} et~al.}{2018}]{ken+18}
{Kennedy} M.~R.,  {Clark} C.~J.,  {Voisin} G.,   {Breton} R.~P.,  2018, \mn@doi
  [\mnras] {10.1093/mnras/sty731}, \href
  {https://ui.adsabs.harvard.edu/abs/2018MNRAS.477.1120K} {477, 1120}

\bibitem[\protect\citeauthoryear{{Linares}}{{Linares}}{2014}]{lin14}
{Linares} M.,  2014, \mn@doi [\apj] {10.1088/0004-637X/795/1/72}, \href
  {http://adsabs.harvard.edu/abs/2014ApJ...795...72L} {795, 72}

\bibitem[\protect\citeauthoryear{{Oh} et~al.,}{{Oh} et~al.}{2018}]{bat105}
{Oh} K.,  et~al., 2018, \mn@doi [\apjs] {10.3847/1538-4365/aaa7fd}, \href
  {https://ui.adsabs.harvard.edu/abs/2018ApJS..235....4O} {235, 4}


\bibitem[\protect\citeauthoryear{{Papitto} et~al.,}{{Papitto}
  et~al.}{2018}]{pap+18}
{Papitto} A.,  et~al., 2018, \mn@doi [\apjl] {10.3847/2041-8213/aabee9}, \href
  {https://ui.adsabs.harvard.edu/abs/2018ApJ...858L..12P} {858, L12}


\bibitem[\protect\citeauthoryear{{Patruno} et~al.,}{{Patruno}
  et~al.}{2014}]{pat+14}
{Patruno} A.,  et~al., 2014, \mn@doi [\apjl] {10.1088/2041-8205/781/1/L3},
  \href {http://adsabs.harvard.edu/abs/2014ApJ...781L...3P} {781, L3}

\bibitem[\protect\citeauthoryear{{Rau} et~al.,}{{Rau} et~al.}{2009}]{ptf09}
{Rau} A.,  et~al., 2009, \mn@doi [\pasp] {10.1086/605911}, \href
  {https://ui.adsabs.harvard.edu/abs/2009PASP..121.1334R} {121, 1334}

\bibitem[\protect\citeauthoryear{{Roberts}}{{Roberts}}{2013}]{rob13}
{Roberts} M.~S.~E.,  2013, in {van Leeuwen} J.,  ed.,  IAU Symposium Vol. 291,
  IAU Symposium. pp 127--132 (\mn@eprint {arXiv} {1210.6903}),
  \mn@doi{10.1017/S174392131202337X}

\bibitem[\protect\citeauthoryear{{Romani} \& {Sanchez}}{{Romani} \&
  {Sanchez}}{2016}]{rs16}
{Romani} R.~W.,  {Sanchez} N.,  2016, \mn@doi [\apj]
  {10.3847/0004-637X/828/1/7}, \href
  {https://ui.adsabs.harvard.edu/abs/2016ApJ...828....7R} {828, 7}

\bibitem[\protect\citeauthoryear{{Romani}, {Filippenko}  \& {Cenko}}{{Romani}
  et~al.}{2015}]{rfc15}
{Romani} R.~W.,  {Filippenko} A.~V.,   {Cenko} S.~B.,  2015, \mn@doi [\apj]
  {10.1088/0004-637X/804/2/115}, \href
  {https://ui.adsabs.harvard.edu/abs/2015ApJ...804..115R} {804, 115}

\bibitem[\protect\citeauthoryear{{Sabbi}, {Gratton}, {Ferraro}, {Bragaglia},
  {Possenti}, {D'Amico}  \& {Camilo}}{{Sabbi} et~al.}{2003}]{sab+03}
{Sabbi} E.,  {Gratton} R.,  {Ferraro} F.~R.,  {Bragaglia} A.,  {Possenti} A.,
  {D'Amico} N.,   {Camilo} F.,  2003, \mn@doi [\apjl] {10.1086/375729}, \href
  {https://ui.adsabs.harvard.edu/abs/2003ApJ...589L..41S} {589, L41}


\bibitem[\protect\citeauthoryear{{Shahbaz} et~al.,}{{Shahbaz}
  et~al.}{2015}]{sha+15}
{Shahbaz} T.,  et~al., 2015, \mn@doi [\mnras] {10.1093/mnras/stv1686}, \href
  {https://ui.adsabs.harvard.edu/abs/2015MNRAS.453.3461S} {453, 3461}

\bibitem[\protect\citeauthoryear{{Shahbaz}, {Dallilar}, {Garner}, {Eikenberry},
  {Veledina}  \& {Gandhi}}{{Shahbaz} et~al.}{2018}]{sha+18}
{Shahbaz} T.,  {Dallilar} Y.,  {Garner} A.,  {Eikenberry} S.,  {Veledina} A.,
  {Gandhi} P.,  2018, \mn@doi [\mnras] {10.1093/mnras/sty562}, \href
  {https://ui.adsabs.harvard.edu/abs/2018MNRAS.477..566S} {477, 566}

\bibitem[\protect\citeauthoryear{{Stappers} et~al.,}{{Stappers}
  et~al.}{2014}]{sta+14}
{Stappers} B.~W.,  et~al., 2014, \mn@doi [\apj] {10.1088/0004-637X/790/1/39},
  \href {http://adsabs.harvard.edu/abs/2014ApJ...790...39S} {790, 39}

\bibitem[\protect\citeauthoryear{{Strader}, {Li}, {Chomiuk}, {Heinke}, {Udalski}, {Peacock}, {Shishkovsky}  \& {Tremou}}{{Strader} et~al.}{2016a}]{slc+16} 
{Strader} J.,  {Li} K.-L.,  {Chomiuk} L.,  {Heinke} C.~O.,  {Udalski} A., 
{Peacock} M.,  {Shishkovsky} L.,   {Tremou} E.,  2016, 
\mn@doi [\apj] {10.3847/0004-637X/831/1/89}, 
\href {http://adsabs.harvard.edu/abs/2016ApJ...831...89S} {831, 89}

\bibitem[\protect\citeauthoryear{{Strader} et~al.,}{{Strader}
  et~al.}{2019}]{str+19}
{Strader} J.,  et~al., 2019, \mn@doi [\apj] {10.3847/1538-4357/aafbaa}, \href
  {http://adsabs.harvard.edu/abs/2019ApJ...872...42S} {872, 42}

\bibitem[\protect\citeauthoryear{{Swihart} et~al.,}{{Swihart}
  et~al.}{2018}]{swi+18}
{Swihart} S.~J.,  et~al., 2018, \mn@doi [\apj] {10.3847/1538-4357/aadcab},
  \href {https://ui.adsabs.harvard.edu/abs/2018ApJ...866...83S} {866, 83}

\bibitem[\protect\citeauthoryear{{Takata} et~al.,}{{Takata}
  et~al.}{2014}]{tak+14}
{Takata} J.,  et~al., 2014, \mn@doi [\apj] {10.1088/0004-637X/785/2/131}, \href
  {http://adsabs.harvard.edu/abs/2014ApJ...785..131T} {785, 131}

\bibitem[\protect\citeauthoryear{{The Fermi-LAT collaboration}}{{The Fermi-LAT
  collaboration}}{2019}]{4fgl19}
{The Fermi-LAT collaboration} 2019, arXiv e-prints, \href
  {https://ui.adsabs.harvard.edu/abs/2019arXiv190210045T} {p. arXiv:1902.10045}

\bibitem[\protect\citeauthoryear{{Torres}, {Ji}, {Li}, {Papitto}, {Rea}, {de
  O{\~n}a Wilhelmi}  \& {Zhang}}{{Torres} et~al.}{2017}]{tor+17}
{Torres} D.~F.,  {Ji} L.,  {Li} J.,  {Papitto} A.~r.,  {Rea} N.,  {de O{\~n}a
  Wilhelmi} E.,   {Zhang} S.,  2017, \mn@doi [\apj]
  {10.3847/1538-4357/836/1/68}, \href
  {https://ui.adsabs.harvard.edu/abs/2017ApJ...836...68T} {836, 68}

\bibitem[\protect\citeauthoryear{{Wang}, {Archibald}, {Thorstensen}, {Kaspi},
  {Lorimer}, {Stairs}  \& {Ransom}}{{Wang} et~al.}{2009}]{wan+09}
{Wang} Z.,  {Archibald} A.~M.,  {Thorstensen} J.~R.,  {Kaspi} V.~M.,  {Lorimer}
  D.~R.,  {Stairs} I.,   {Ransom} S.~M.,  2009, \mn@doi [\apj]
  {10.1088/0004-637X/703/2/2017}, \href
  {http://adsabs.harvard.edu/abs/2009ApJ...703.2017W} {703, 2017}

\bibitem[\protect\citeauthoryear{{Xing} \& {Wang}}{{Xing} \&
  {Wang}}{2015}]{xw-XSS}
{Xing} Y.,  {Wang} Z.,  2015, \mn@doi [\apj] {10.1088/0004-637X/808/1/17},
  \href {https://ui.adsabs.harvard.edu/abs/2015ApJ...808...17X} {808, 17}

\bibitem[\protect\citeauthoryear{{Xing} \& {Wang}}{{Xing} \&
  {Wang}}{2016}]{xw16}
{Xing} Y.,  {Wang} Z.,  2016, \mn@doi [\apj] {10.3847/0004-637X/831/2/143},
  \href {https://ui.adsabs.harvard.edu/abs/2016ApJ...831..143X} {831, 143}

\bibitem[\protect\citeauthoryear{{Xing}, {Wang}  \& {Takata}}{{Xing}
  et~al.}{2018}]{xwt18}
{Xing} Y.,  {Wang} Z.-X.,   {Takata} J.,  2018, \mn@doi [Research in Astronomy
  and Astrophysics] {10.1088/1674-4527/18/10/127}, \href
  {https://ui.adsabs.harvard.edu/abs/2018RAA....18..127X} {18, 127}

\bibitem[\protect\citeauthoryear{{Yap}, {Li}, {Kong}, {Takata}, {Lee}  \&
  {Hui}}{{Yap} et~al.}{2019}]{yap+19}
{Yap} Y.~X.,  {Li} K.~L.,  {Kong} A.~K.~H.,  {Takata} J.,  {Lee} J.,   {Hui}
  C.~Y.,  2019, \mn@doi [\aap] {10.1051/0004-6361/201834545}, \href
  {https://ui.adsabs.harvard.edu/abs/2019A&A...621L...9Y} {621, L9}

\bibitem[\protect\citeauthoryear{{van Paradijs} \& {McClintock}}{{van Paradijs}
  \& {McClintock}}{1995}]{vm95}
{van Paradijs} J.,  {McClintock} J.~E.,  1995, in X-ray Binaries. pp 58--125

\makeatother
\end{thebibliography}


\newpage

\begin{landscape}
\begin{table}
\centering
\caption{\fermi\ LAT flux measurements of 4FGL J0935.3$+$0901}
\label{tab:spectra}
\begin{tabular}{lccccccc}
\hline
\multicolumn{2}{c}{ } &
\multicolumn{2}{c}{Total period} &
\multicolumn{2}{c}{High-flux period} &
\multicolumn{2}{c}{Low-flux period} \\ \hline
$E$ & Band & $F/10^{-12}$ & TS & $F/10^{-12}$ & TS & $F/10^{-12}$ & TS \\
(GeV) & (GeV) & (erg cm$^{-2}$ s$^{-1}$) & & (erg cm$^{-2}$ s$^{-1}$) & & (erg c
m$^{-2}$ s$^{-1}$) & \\ \hline
0.15 & 0.1--0.2 & 1.2$\pm$0.6 & 10 & 4$\pm$1 & 20 & $<$1.1 & 0 \\
0.36 & 0.2--0.5 & 1.5$\pm$0.2 & 48 & 2.0$\pm$0.5 & 21 & 1.2$\pm$0.3 & 21 \\
0.84 & 0.5--1.3 & 1.2$\pm$0.2 & 84 & 1.6$\pm$0.3 & 37 & 0.9$\pm$0.2 & 32 \\
1.97 & 1.3--3.0 & 0.9$\pm$0.1 & 73 & 1.0$\pm$0.3 & 19 & 0.9$\pm$0.2 & 49 \\
4.62 & 3.0--7.1 & 0.3$\pm$0.1 & 12 & 0.5$\pm$0.2 & 6 & 0.2$\pm$0.1 & 6 \\
10.83 & 7.1--16.6 & $<$0.3 & 2 & $<$0.9 & 1 & $<$0.2 & 0 \\
25.37 & 16.6--38.8 & $<$0.4 & 0 & $<$0.6 & 0 & $<$0.8 & 0 \\
59.46 & 38.8--91.0 & $<$0.5 & 0 & $<$1.4 & 0 & $<$0.9 & 0 \\
139.36 & 91.0--213.3 & $<$0.9 & 0 & $<$3.2 & 0 & $<$1.4 & 0 \\
326.60 & 213.3--500.0 & $<$2.0 & 0 & $<$7.7 & 0 & $<$3.3 & 0 \\
\hline
\end{tabular}
\vskip 1mm
\footnotesize{Note: $F$ is the energy flux ($E^{2} dN/dE$).  
Upper limits are quoted at a confidence level of 95\%.  }
\end{table}
\end{landscape}

\bsp	
\label{lastpage}
\end{document}